# Accuracy Enhancement of an Optical Network Digital Twin Based on Open-Source Field Data

Ambashri Purkayastha [1,3], Camille Delezoide[1], Mounia Lourdiane[2], Cédric Ware[3], and Patricia Layec[1]

[1] Nokia Bell Labs, France, ambashri.purkayastha@nokia.com
[2] SAMOVAR, Télécom SudParis, Institut Polytechnique de Paris, France
[3] LTCI, Télécom Paris, Institut Polytechnique de Paris, France

**Abstract** *We propose a two-stage hybrid QoT model for twinning a real transport network and evaluate it on recently published field data. Accounting for partial calibration of key parameters, we improve the SNR prediction accuracy by more than a factor of two. ©2024 The Author(s)*

## Introduction

Quality of transmission (QoT) estimation is of strategic importance for designing and upgrading optical networks with just enough margins to meet service level agreements during an optical network's lifetime [1].

Among the many QoT estimation models proposed recently [2], hybrid models combining both physics-derived equations and monitoring data - to refine parameter knowledge and/or the model itself - should achieve the best trade-off between generalizability and accuracy.

Recently, hybrid QoT estimation converged with the concept of digital twin (DT) [3-5]. The DT concept is particularly powerful for predicting the impact of any network modification on existing and new lightpaths, or to test the general resilience of a network when simulating failures [6]. Hybrid QoT estimation is particularly suited at the core of a network digital twin: monitoring data can be directly leveraged to enhance a general physics-based QoT model at all time, thus ensuring the fidelity of the DT to its physical counterpart.

Many hybrid QoT models have been proposed in the literature and tested in simulations and experiments. Yet, the industry-wide adoption of such models has been hindered by the inherent difficulty to properly assess and compare their performances in deployed networks, where multiple deviations from typical model assumptions are often observed [7-9]. However, such assessment has now been made possible by the recent publication of a dataset fully described in [10,11]. It reports the QoT of 25 lightpaths from four distinct groups, as well as all input and output powers at the amplifiers, and channels powers measured after each amplifier by optical channel monitors (OCM), all monitored hourly over two weeks. The dataset also contains essential calibration data, further discussed in the paper.

In the following, we will use the QoT estimation model presented and assessed on the same dataset in [12]—that we refer to as (M$_1$)—as performance baseline. We further propose a two-stage hybrid QoT model that progressively enhances the estimation accuracy over time by correcting for both static and time-dependent parameters, separately. We finally validate the method over the open-source dataset and compare its performance with the baseline model (M$_1$).

## Transmission Modelling

The principal QoT metric in optical networks is the bit error ratio (BER) before forward error correction (FEC). This pre-FEC BER can be directly converted into a signal-to-noise ratio (SNR) [13].

This SNR primarily depends on three types of noise: the amplified spontaneous emission (ASE) noise from the optical amplifiers, the nonlinear interference (NLI) due to the Kerr effects in the fiber, and the so-called transponder (TRX) noise due to various performance-limiting effect from both transmitter and receiver. In contrast, the generalized OSNR (GOSNR) characterizes the performance of the line independently from that of the transponder. It is generally defined as:

$$\text{GOSNR} = \frac{P_s}{P_{ASE} + P_{NLI}} \qquad (1)$$

where $P_s$ is the total optical signal power, while $P_{ASE}$ and $P_{NLI}$ are respectively the total ASE and NLI noise powers in a 0.1-nm bandwidth $B_o$. All three quantities are measured just before the receiver.

The ASE power for the k$^{th}$ channel in the n$^{th}$ amplifier is calculated from the gain $G_{n,k}$ and noise figure $NF_{n,k}$ as: $P_{ASE_{n,k}} = NF_{n,k}\, G_{n,k} h\nu_k B_{ref}$; where $B_{ref}$ is the reference bandwidth, and $\nu_k$ is the channel's center frequency. The total ASE power at the end of transmission is calculated as the sum of ASE powers generated per device as $P_{ASE} = \sum P_{ASE_n}$

For the nonlinear noise, we leverage a closed-form version of the IGN model in [14] to estimate the power of NLI for the j$^{th}$ span ($P_{NLI_j}$). The total NLI accumulated at the end of the transmission is

$P_{NLI} = \sum P_{NLI_j}$. NLI and ASE powers are first calculated in the $B_{ref}$, and then converted to the $B_o$ for calculating the GOSNR.

From back-to-back calibration available in the dataset, we deduce the relation between SNR and GOSNR for DP-QPSK modulation. The SNR-GOSNR curve displayed in Fig. 1 is critical for comparing measured SNR values $SNR_{meas}$ to the GOSNR estimated values $GOSNR_{est}$ from the ASE and NLI models. $GOSNR_{int}$ is defined as the GOSNR interpolated from $SNR_{meas}$ from the curve in Fig. 1.

## QoT Estimation for Digital Twin

To evaluate the accuracy gain achieved when monitoring is leveraged to assess model parameters, we reproduce a network design model $M_0$ by considering the worst-case value for all parameters. Upon deployment, we incorporate the measured values of the parameters used for QoT estimation, resulting in model $M_1$. In this paper, we refine the parameters observed in $M_1$ by correcting measurement biases, and accounting for missing frequency and power dependent calibrations which we will refer to as $M_2$.

The model $M_1$ in [12] is meant to provide $SNR_{est}$ values as soon as the lightpath under study is active. It leverages all monitoring data available in the field dataset. The total launch power $(P_s)$, into the fiber is calculated from the channel powers $(P_{s_k})$ measured at the OCMs and the card insertion loss $(l_{IL})$ from [11] as:

$$P_s = \sum P_{s_k} - l_{IL} \quad (2)$$

The gain in dB for the $k^{th}$ channel of the $n^{th}$ inline EDFA is calculated as:

$$G_{n,k} = P_{s_k,n} - (P_{s_k,n-1} - l_{agg}) \quad (3)$$

Where, $l_{agg}$ accounts for the fiber loss and all insertion losses before the signal enters the EDFA. Here, we then further calculate $NF_{n,k}$ as a function of $G_{n,k}$ by interpolating it from the G-NF calibration data in [11]. The estimation error at time t is then calculated as:

$$\varepsilon(t) = GOSNR_{est,t} - GOSNR_{int,t} \quad (4)$$

In practice, independently from the accuracy of the ASE and NLI models, there are multiple explanations for any systematic offset between $GOSNR_{int}$ and $GOSNR_{est}$ and by extension, $SNR_{est}$ and $SNR_{meas}$. From [12], we know that OCMs are prone to measurement errors. The G-NF curve assumes that the gain and NF of an EDFA is frequency independent, but this is contradicted by experiments and simulations [15,16,17]. Interpolation of $GOSNR_{int}$ from $SNR_{meas}$ assumes that the calibration is accurate regardless of the individual transponder card, filter cascade, channel center frequency and receiver's input power. However, those parameters can have a strong impact on the calibration curve, especially the receiver's input power $(P_{rx})$. [18]

Thus, we propose model $M_2$ to correct key parameters once the lightpath under study has been observed for sufficient time. This can be achieved in two stages: in $M_{2a}$, we correct the median offset in the GOSNR, and in $M_{2b}$, the amplitude of SNR variations.

Firstly, assuming there exists a time-independent, frequency-dependent measurement bias at the OCMs and EDFA, we correct the channel power and NF to minimize the root-mean-square (rms) error. Over a small training window of time $(T_{train})$, we calculate $\vartheta_k$ as the median of $\varepsilon_k(T_{train})$, and add $\vartheta_k$ to $P_{s_k}$ to correct for measurement bias. After accounting for this correction on the power, if $\varepsilon'(T_{train}) \not\approx 0$, we recalculate $\vartheta'_k$ from $\varepsilon'(T_{train})$, and subtract $\vartheta'_k$ from $NF_{n,k}$. This allows us to correct the constant error in $M_{2a}$.

Secondly, we observe from the dataset that all transmissions are short distance transmissions. Performance is thus heavily limited by transponder noises, which directly impact the SNR-GOSNR curve through the $SNR_{trx}$, as illustrated in Fig. 1. Therefore, not accounting for the variations of this parameter can lead to a strong time dependent error. From [18], we know that the $SNR_{trx}$ can have a non-linear dependence on the receiver power $P_{rx}$, which should be subject to further calibration for each transponder. Since we do not have the direct channel power measurement at the receiver, in $M_{2b}$, we assume that the $\Delta P_{rx}$ is proportional to $\Delta P_{ocm,last}$. Thus accounting for this evolution of the channel powers at the last OCM will partially account for the variation of $SNR_{trx}$. In the training window $T_{train}$, we perform a linear regression between $SNR_{meas}$ and $P_{ocm,last}$, such that: $\Delta SNR = \theta \cdot \Delta P_{ocm,last} + SNR_0$, where $\theta$ is the slope of the linear fit, and $SNR_0$ is the measured SNR when $P_{ocm}=0$. Then we can write $SNR_{est} = SNR_{int} + \Delta SNR$, where $SNR_{int}$ is the interpolated SNR of the $GOSNR_{est}$ from the calibration curve in Fig. 1. This gives us a model that accounts for the static power and NF offsets, as well as the time variant $SNR_{trx}$.

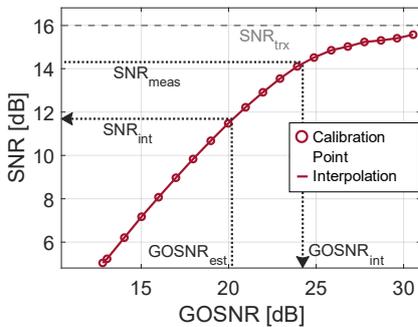

**Fig. 1:** Calibration curve of GOSNR vs SNR derived from GOSNR-BER calibration in the dataset for optical transponder 1.

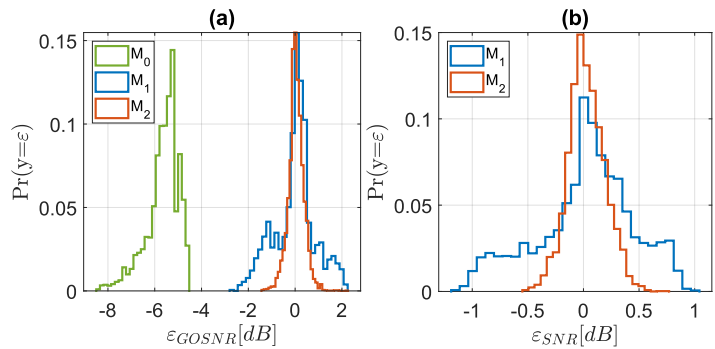

**Fig. 2:** Comparison of accuracy at design with $M_0$, after deployment with availability of monitoring data with $M_1$, and with parameter refinement with $M_2$ for (a) GOSNR, (b) SNR.

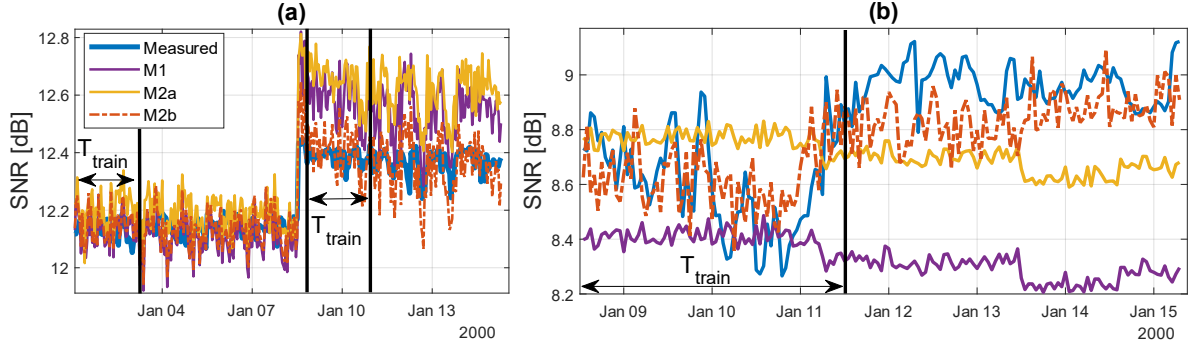

**Fig. 3:** Time evolution of SNR estimations for two example lightpaths (a) Group 2, Channel 1; (b) Group 3, Channel 1

**Results**

In Fig. 2, we present the distributions of GOSNR error $\varepsilon_{GOSNR}$, from $M_0$, $M_1$, and $M_2$ and SNR error $\varepsilon_{SNR}$ by $M_1$, and $M_2$ over all lightpaths. In Fig. 3 we illustrate the time evolution of SNR estimation we attain per model for two example channels on separate lightpaths. Fig. 3a presents a case where the EDFA settings were deliberately changed on Jan 8th for the entire lightpath. We retrain the model to re-estimate parameters with updated network configuration. Tab. 1 gives the rms error over all channels and unique lightpath groups calculated from all three models showing progressive improvement in estimation.

In Fig. 2, $\varepsilon_{GOSNR}$ has a larger variance than $\varepsilon_{SNR}$, since the ASE and NLI noise are masked by the transponder noise. In estimation using $M_0$, there is an offset of the mean error by $-6$ dB in GOSNR, and $-4$ dB in SNR. This underestimation is a deliberate result of considering the worst of all parameters at design. Accounting for the monitoring data in $M_1$ corrects the offset in mean error, and the further correction of parameters in $M_2$ reduces the variance of the error distribution.

In Fig. 3a, before Jan 8, $M_1$ predicts the performance within 0.086 dB rms error. Although it predicts the network state change, the rms error increases to 0.26 dB. In case of Fig. 3b, the first estimation using monitoring data results in rms error of 0.49 dB, due to incomplete knowledge of channel power, NF, and $SNR_{trx}$.

In Fig. 3b when we compare the time-dependent estimations of the SNR from $M_1$ and $M_2$, we show an initial improvement by correcting the median offset in $M_{2a}$, and then a further refinement by accounting for the $SNR_{trx}$ evolution in time in $M_{2b}$. In $M_{2a}$, the $\varepsilon_{SNR}$ is first minimized in the $T_{train}$. The performance is then predicted for the remaining transmission time. In Fig. 3a, after the state change, $M_{2a}$ accounts for the change of EDFA settings but fails to account for the change of $SNR_{trx}$, resulting in rms error of 0.2 dB. By retraining the system, and recalibrating the $SNR_{trx}$ with $M_{2b}$ in the $T_{train}$, we reduce the estimation rms error to 0.12 dB.

The overall rms error ($G_T$) over all lightpaths in all groups is reduced from 0.43 dB by monitoring data to 0.17 dB by our 2-stage parameter correction method.

**Conclusion**

We show the advantage of using a hybrid DT for QoT estimation that learns from live measurements for a short period of time, corrects insufficient calibration data, and predicts the performance for the remaining transmission time with this correction, halving the rms error, with the limitation that this final model is only valid locally, i.e. for the lightpath under study. We demonstrate the importance of monitoring the receiver power, especially at short distance transmissions, where the transponder noise is dominant to further enhance the accuracy of QoT estimation. We also show the importance of extensive power and frequency calibration of key parameters like SNR of the transponders, and the amplifier noise figure.

|    | G1   | G2   | G3   | G4   | $G_T$ |
|----|------|------|------|------|------|
| M0 | 4.36 | 4.33 | 4.46 | 4.17 | 4.36 |
| M1 | 0.43 | 0.15 | 0.52 | 0.45 | 0.43 |
| M2 | 0.18 | 0.12 | 0.20 | 0.17 | 0.17 |

**Tab. 1:** RMS error (dB) over each and all lightpath groups

**Acknowledgements**

This work has been supported by the French government through the Celtic-Next AINET-ANTILLAS research project.